\documentclass[preprintnumbers,twocolumn,aps,prd,amsfonts,amsmath,amssymb,nofootinbib,longbibliography]{revtex4-1}

\usepackage{braket}

\usepackage{ifpdf}
\ifpdf
    \usepackage[pdftex,colorlinks=true,plainpages=false,pdfpagelabels]{hyperref}
    \usepackage[pdftex]{color}
    \usepackage[pdftex]{graphicx}
    \hypersetup{colorlinks = true}
\else
    \usepackage[colorlinks=true]{hyperref}
    \usepackage{graphicx}
\fi

\def\half{\frac{1}{2}}

\begin{document}

\title{Quantum Instability of the Emergent Universe}
\preprint{IPMU13-0121}

\author{Anthony Aguirre}
\email{aguirre@scipp.ucsc.edu}
\affiliation{Santa Cruz Institute for Particle Physics and Department of Physics, University of California, Santa Cruz, CA 95064, USA}
\author{John Kehayias}
\email{john.kehayias@ipmu.jp}
\affiliation{Kavli Institute for the Physics and Mathematics of the Universe (WPI), Todai Institutes for Advanced Study, The University of Tokyo, Kashiwa, Chiba 277-8583, Japan}

\date{\today}

\begin{abstract}
  \noindent
  We perform a semi-classical analysis of the Emergent Universe
  scenario for inflation. Fixing the background, and taking the
  inflaton to be homogenous, we cast the inflaton's evolution as a
  one-dimensional quantum mechanics problem. The potential is taken to
  be flat or linear, as an approximation to the monotonic and
  slowly-varying asymptotic behavior of the Emergent Universe
  potential. We find that the tuning required over a long time scale
  for this inflationary scenario is unstable quantum
  mechanically. Considering the inflaton field value as a wavepacket,
  the spreading of the wavepacket destroys any chance of both starting
  and ending with a well-formed state. Thus, one cannot have an
  Einstein static universe to begin with that evolves into a
  well-defined beginning to inflation a long time later.
\end{abstract}

\maketitle

\section{Introduction}

The ancient question of whether the Universe had a beginning, or has
existed eternally, has in recent decades been brought into focus using
the tools and knowledge of General Relativity (GR) and modern
cosmology. The question's resolution, however, is still far from
clear.  The standard Friedmann-Lema\'itre-Robertson-Walker (FLRW)
big-bang cosmology includes an initial singularity, and there is a
number of theorems purporting that such singularities are generic
\cite{singularity_theorem1, *singularity_theorem2,
  *singularity_theorem3, *singularity_theorem4, *singularity_theorem5,
  *singularity_theorem6}, suggesting that \emph{classical} spacetime,
at least, has a ``beginning'' in the sense of a global spacelike
surface at which classical GR breaks down.  However, these theorems
all brook various exceptions and loopholes, and several scenarios have
been developed that circumvent these theorems and form the basis for
classical or semi-classical ``past eternal'' cosmologies (see
\cite{Aguirre:2007gy} for review of some of these involving
inflation).

One scenario that has garnered significant attention is the ``Emergent
Universe'' of Ellis \& Maartens \cite{emergent_universe,
  *Ellis:2003qz}.  This cosmology has several attractive features:
there is no initial singularity or ``beginning of time,'' no horizon
problem, and (it is claimed) no quantum gravity era because the
curvature scale always greatly exceeds the Planck scale. In this
scenario at some early time the Universe is a closed FLRW cosmology
that as time $t\rightarrow -\infty$ asymptotes to an Einstein static
universe, with a negative-pressure energy component that stabilizes
the universe against gravitational collapse.  Thus the closed universe
exists ``eternally,'' but then at some point begins inflation. That
is, the first e-folding of inflation take an unbounded amount of time,
but the second and subsequent e-foldings proceed essentially as usual,
ending in reheating and ordinary cosmological evolution.

A key question about this scenario is whether the initial state can
self-consistently exist for eternity.  Classically, the Einstein
static universe is unstable to homogeneous perturbations, but stable
to inhomogeneous perturbations if the fluid sound speed is
sufficiently high \cite{es_stability0, *es_stability1, *es_stability2,
  *es_stability3, *es_stability4}.  This indicates that eternality is
possible in principle, but only if the homogeneous mode of the
positive-pressure content \emph{precisely} balances the negative
energy repulsive component of the energy density.  This appears
problematic, however, in that we might expect that quantum
fluctuations can destabilize this careful balance, causing the
universe to collapse or expand uncontrollably.

In this note, we examine the simplest version of the Emergent
Universe, based on a single rolling scalar field in GR.  Our analysis
will treat the metric classically (and in fact fixed), and assume the
scalar field is homogeneous, but quantized.  Treating the metric
quantum mechanically (e.g.~using the Wheeler-deWitt formalism) or
including anisotropic perturbations seems very unlikely to increase
the cosmology's stability.  These assumptions allow us to cast the
problem as a simple 1D quantum mechanics problem, following the method
of \cite{infl_qm1, *infl_qm2}, where the degree of freedom is the
value of the scalar field. This reveals a result that is intuitively
perhaps unsurprising: due to the spreading of any wavepacket, it is
inconsistent to have a well-defined state (Gaussian wavepacket) in the
asymptotic past as an Einstein static universe, as well as a
well-defined state at the beginning of inflation (i.e.~at the end of
the first e-folding). We show this by analyzing both a flat and nearly
flat (linear) potential for the wavepacket evolution. Computing the
probability of the initial state evolving into a well-defined
pre-inflationary state, we see that it is severely suppressed (and
effectively zero in the `past eternal' limit). We discuss these
calculations and also comment on the Emergent Universe in alternative
gravitational theories.

\section{The Emergent Universe}

The Emergent Universe evades relevant singularity theorems
\cite{singularity_theorem1, *singularity_theorem2,
  *singularity_theorem3, *singularity_theorem4, *singularity_theorem5,
  *singularity_theorem6} by explicitly violating the assumptions that
$K = -1, 0$ and $H \equiv \dot{a}/a > 0$. The Emergent Universe is
closed, $K = +1$, and has $H = 0$ initially. This scenario does not
bounce, but starts as an Einstein static universe with finite size in
the infinite past, inflates, and then reheats in the usual manner. In
this way it avoids an initial singularity and horizon problem, and
the initial size can be large enough for this scenario to avoid a
quantum gravity era. Although there is an infinite time for inflation,
the amount of inflation is finite (and can be made large).

The simplest setup for the Emergent Universe is an Einstein static
universe with a cosmological constant (absorbed into the constant term
of the scalar field's potential, $V(\phi)$) and minimally coupled
scalar field, $\phi$. The Friedmann equations are
\begin{align}
\frac{\ddot{a}}{a} = -\frac{8\pi G}{3}\left(\dot{\phi}^2 - V(\phi)\right),\label{eq:fried1}\\
H^2 = \frac{8\pi G}{3}\left(\half\dot{\phi}^2 + V(\phi)\right) - \frac{K}{a^2},
\end{align}
where a dot represents a time derivative, $G$ is Newton's constant,
and $a(t)$ is the scale factor using the conventions
of~\cite{emergent_universe, *Ellis:2003qz,infl_qm1} (it has units of
length\footnote{The mass dimension of other quantities is $[\phi] = 1,
  [\sigma_0] = 1, [T] = 2$.}).

A positive minimum for the (initial) scale factor, $a_0$,\footnote{The
  dominance of the curvature term in the past, for sufficiently long
  inflation, allows such a solution with $K = +1$
  \cite{Ellis:2001ym}.} has $H_0 = 0$ and
\begin{equation}
\frac{3K}{8\pi G a_0^2} = \half\dot{\phi_0}^2 + V_0,
\end{equation}
where the zero subscript denotes the initial value (the same time as
$a_0$ is defined). Furthermore eq.~\eqref{eq:fried1}, since $\ddot{a}
= 0$, tells us that
\begin{equation}
\dot{\phi_0}^2 = V_0.
\end{equation}
Then the value of the potential at this minimum, or the initial vacuum
energy, is
\begin{equation}
V_0 = \frac{K}{4\pi G a_0^2} (= \dot{\phi}_0^2).
\end{equation}

Here we see the beginning of a potential problem for the Emergent
Universe: the Einstein static universe requires a precise balancing of
the kinetic energy of the scalar field with the vacuum energy. This
balancing must persist if the universe is to be considered static and
thus will be sensitive to quantum fluctuations in the scalar field. In
this note we attempt to analyze this problem via a
``semi-classical''-like analysis.

One might also worry about the classical stability of the Einstein
static universe \cite{es_stability0, *es_stability1, *es_stability2,
  *es_stability3, *es_stability4}. For the simplest case we are
considering, where there is no matter, the static universe is
neutrally stable for inhomogeneous linear perturbations. Homogeneous
perturbations will break the balance of the curvature to vacuum
energy, leading to inflation, thus to perdure for an indefinite amount
of time, this balance in the zero-mode must be mathematically perfect.
We do not address concerns about this here.\footnote{For example, it
  seems likely that nonlinear coupling between the modes would leak
  power from inhomogeneous modes to the homogeneous mode, making the
  universe effectively unstable to all perturbations
  \cite{Losic:2004hw}.} Rather, we assume that such a perfect balance
is maintainable classically, and investigate the same problem when
quantized.

\section{Inflation as One-Dimensional Quantum Mechanics}

First, let us define our setup and conventions, which follows closely
from \cite{emergent_universe, *Ellis:2003qz, infl_qm1}. We will take
the simplest case of the universe filled with just a scalar field,
$\phi$, in a FLRW background with $K=+1$, scale factor $a(t)$, and
Hubble expansion rate $H\equiv \dot{a}/a$. The scalar field obeys
\begin{equation}
  \ddot{\phi} + 3H\dot{\phi} + V'(\phi) = 0,
\end{equation}
where the dots denote time derivatives, $V(\phi)$ is the inflaton
potential, and the prime denotes a derivative with respect to the
argument.

We can cast the evolution of the inflaton as a one-dimensional quantum
mechanics problem by making the following simplifications:
\begin{itemize}
\item Treat the background geometry classically (and fixed); the scale
  factor is not treated as a field.
\item Take the inflaton to be homogeneous and its value everywhere in
  space, $\phi$, to be the ``coordinate.''
\item Take the inflaton momentum to also be homogeneous, and its
  value, $\dot{\phi}$, is proportional to the conjugate momentum.
\end{itemize}

From this setup, one sees that we are performing a type of ``semi-''
or ``quasi-classical'' analysis. For the Emergent Universe scenario in
particular, this type of analysis will shed light on the question of
stability beyond purely classical considerations by being the next
step in sophistication. The Emergent Universe's behavior in the
asymptotic past should be that of a static universe, which then evolves
ever so slowly for an infinite amount of time. It makes sense then to
treat the background classically and independently from the scalar
field. However, the delicate tuning of the scalar field's kinetic
energy leads us to consider any small deviations, especially over the
infinite amount of time. Thus, we treat the scalar field in a quantum
mechanical manner. And, as we shall see in the following sections, our
setup is enough to see a serious instability or inconsistency in the
Emergent Universe's evolution.

The Lagrangian\footnote{Note that this is not a Lagrangian
  \emph{density}; we've already integrated over space, hence the
  volume factor of $2\pi a^3(t)$.} is
\begin{equation}
  L = 2\pi a^3(t) \left(\half\dot{\phi}^2 - V(\phi)\right).
\end{equation}
After performing the usual Legendre transformation, with conjugate
momentum $p \equiv 2\pi^2a^3(t)\dot{\phi}$, the Hamiltonian is
\begin{equation}
  H = \half\frac{p^2}{2\pi^2a^3(t)} + 2\pi^2a^3(t)V(\phi).
\end{equation}

We now define our ``wavefunction'' as $\psi(\phi,t)$ which satisfies
the Schr\"odinger equation from the above Hamiltonian
\begin{equation}
  i\frac{\partial\psi}{\partial t} =
  -\half\frac{1}{2\pi^2a^3}\frac{\partial^2\psi}{\partial\phi^2} + 2\pi^2a^3V(\phi)\psi.
\end{equation}
Taking the potential to have at most quadratic terms and after the
following (``conformal time''-like) time coordinate change,
\begin{equation}
  T = \int\frac{1}{a^3(t')}\mathrm{d}t',
\end{equation}
we rewrite the Schr\"odinger equation as
\begin{equation}
  i\frac{\partial\psi}{\partial T} =
  -\frac{1}{4\pi^2}\frac{\partial^2\psi}{\partial\phi^2} + u(\phi,T)\psi,
\end{equation}
with the potential $u(\phi,T)$ given by
\begin{equation}
  u(\phi, T) = 2\pi^2a^6(T)\left(c_2\phi^2 + c_1\phi + c_0\right),
\end{equation}
with the only $T$-dependence coming from the prefactor $2\pi^2a^6(T)$
and the $\phi$-dependence explicit.

We now assume a Gaussian wavepacket form for $\psi$ which we
parametrize as
\begin{equation}
  \psi(\phi, T) = A(T)e^{-B(T)[\phi - f(T)]^2},
\end{equation}
with $A, B,$ and $f$ as arbitrary functions of $T$ to be solved
for. We plug $\psi$ into the Schr\"odinger equation, and by matching
coefficients of each power of $\phi$, we have a set of differential
equations for $A, B,$ and $f$ (for more details see \cite{infl_qm1,
  infl_qm2}).

It will also be useful to think of the wavefunction in momentum space
(the conjugate momentum, as defined above), which is given by the
usual Fourier transform,
\begin{equation}
  \widetilde{\psi}(p, T) = \widetilde{A}(T)e^{-[p + 2iB(T)f(T)]^2/4B(T)},
\end{equation}
with $\widetilde{A}(T) \equiv A(T)\exp[-B(T)f(T)^2]/\sqrt{2}B(T)$.

\section{Instability of the Emergent Universe Scenario}

We can now apply the above framework to the Emergent Universe. We will
model the potential as being composed of sections that are completely
flat, and sections with a constant slope. This could model a
potential that is perfectly flat as $\phi \rightarrow -\infty$,
connected to a sloping portion at $\phi > 0$, or one constructed out
of segments that have a slope approaching zero as $\phi\rightarrow
-\infty$. In either case, we will consider the field at some time
$T=-T_0$ to be in a Gaussian wavepacket centered at $\phi_0$ with
initial spread $\sigma_0$ and moving with velocity $v_0$ (which may be
zero). We then evolve the wavepacket to later times.

Formulated this way, stability concerns arise almost immediately.  For
a given value of $V(\phi)$, only one precise value of $\dot \phi$
yields stability, yet $\phi$ and $p \propto \dot\phi$ are subject to
an uncertainty relation.\footnote{The wavepackets are initially
  minimum uncertainty wavepackets, saturating the uncertainty
  relationship between $\phi$ and $p$, $\sigma_\phi\sigma_p \ge
  1/2$. As the packet evolves in the following potentials, however,
  $\sigma_\phi$ grows while $\sigma_p$ is constant.} In
momentum-space, the wavepacket has nonzero width $\Delta \propto
1/\sigma_0$, so at a given time, unless $\Delta \rightarrow 0$, there
is an infinitesimal probability of a measurement yielding the value of
$\dot\phi$ which gives stability. With probability approaching unity,
the field velocity would have a value for which the universe would
evolve away from the emergent dynamics, into either empty de Sitter
space or a big crunch. Yet in the $\Delta \rightarrow 0$ limit, the
value of $\phi$ is completely uncertain and one cannot describe the
situation as a single classical universe.

Similarly, if we {\em assume} that the universe can always be treated
in a quasi-classical way (as implicitly assumed
by~\cite{emergent_universe, *Ellis:2003qz}), it should have compact
support in both field and field-velocity space. We can then ask: if
the wavefunction at time $-T_0 \rightarrow -\infty$ is a wavepacket of
finite width in both $\phi$ and $\dot\phi$, is there {\em any}
probability that it will evolve into a quasi-classical configuration
at time $T = 0$ at which inflation starts? This can be computed within
our model for either a flat or a constant-slope potential, as detailed
in the next section and in Appendix~\ref{apx:linear}.

\subsection{In a Flat or Linear Potential}\label{sec:const}
As a simplest calculation, we study the evolution of a wavepacket in a
completely flat potential ($c_2 = c_1 = 0$), which mimics the
asymptotic behavior of the Emergent Universe. In this case we will
need to give the initial wavepacket some ``kick'' to have an initial
non-zero velocity.

The initial state (at time $T = -T_0$) is a Gaussian wavepacket
centered at $\phi_0$ with initial spread $\sigma_0$ and moving with
velocity $v_0$:
\begin{equation}
  \psi(\phi, -T_0) =
  \left(\frac{1}{2\pi\sigma_0^2}\right)^{1/4}e^{\frac{-(\phi -
      \phi_0)^2}{4\sigma_0^2} + iv_0\phi_0}.
\end{equation}
At a time $T$ the wavepacket evolves to
\begin{widetext}
\begin{equation}
  \psi(\phi, T) =
  \frac{1}{\left(2\pi\sigma_0^2\right)^{1/4}}\exp\left\{-v_0^2\sigma_0^2 + iv_0\phi_0 - \frac{\pi^2}{\sigma(T)}\left(\phi -\phi_0 - 2iv_0\sigma_0^2\right)^2 -\frac{i}{2}\int_{-T_0}^{T}\frac{1}{\sigma(T')}\left[1 + 2\sigma(T')c(T')\right]\mathrm{d} T'\right\},
\end{equation}
\end{widetext}
with
\begin{equation}\label{eq:sigma}
  \sigma(T) \equiv 4\pi^2\sigma_0^2 + i(T_0 + T),
\end{equation}
and $c(T) \equiv 2\pi^2a^6(T)c_0$. The center of the wavepacket, which is
$\braket{\phi}$, moves with constant velocity $v$ as\footnote{This is
  an example of an expression we rederived that is also in
  \cite{infl_qm1, *infl_qm2} where we differ by a factor of $1/2$ in
  front of $T$ (the other notable instances are the probability
  densities in \cite{infl_qm1, infl_qm2}). We believe this is a typo
  in these papers, but these factors make no difference for our
  analysis.}
\begin{equation}
\braket{\phi} = \phi_0 + \frac{(T_0 + T)v_0}{2\pi^2},
\end{equation}
and the quantum mechanically uncertainty in $\phi$ is
\begin{equation}
\sigma_\phi^2 = |\sigma|^2/16\pi^4\sigma_0^2,
\end{equation}
while the conjugate momentum has
\begin{equation}
\braket{p} = v_0, \qquad \sigma_p = \frac{1}{4\sigma_0}.
\end{equation}

We now want to calculate the probability that a well-formed initial
wavepacket at $T = -T_0$, centered at $\phi = \phi_0$, will evolve
into another ``nice'' wavepacket some time later, $T = 0$, in this
potential. For the final state to compare with we will use the initial
wavepacket with its center shifted to be lined up with the wavepacket
that evolved after time $T_0$ (to time $T = 0$). In other words, the
initial state is $\psi(\phi, -T_0)$ which evolves into $\psi(\phi, 0)$
which we compare to $\psi(\phi, -T_0)|_{\phi_0 = T_0v_0/2\pi^2}$. The
probability we are calculating is
\begin{align}
  P &= \left|\braket{\phi, -T_0\left|_{\phi_0 = T_0v_0/2\pi^2}\big|e^{-iHT_0}\right|\phi,-T_0}\right|^2\nonumber\\
     &= \left|\braket{\phi, -T_0\left|_{\phi_0 = Tv_0/2\pi^2}\right|\phi, 0}\right|^2.
\end{align}

Before specifying the scale factor, the probability is
\begin{equation}
P = \left(\frac{T_0^2}{\chi^2} + 1\right)^{-1/2},
\end{equation}
with $\chi \equiv 8\pi^2\sigma_0^2$. For a static universe the scale
factor is a constant, which we set to $a_0$, and $T = t/a_0^3$. For a
long evolution we take $T_0 \gg \chi$ and the probability, to leading
order in $t$, is
\begin{equation}\label{eq:constp}
P \approx \frac{a_0^3\chi}{t_0} \ll 1.
\end{equation}
Therefore, for a long evolution ($-t_0 \rightarrow -\infty$) in a
constant potential, the probability of a wavepacket evolving into
another well-defined wavepacket after a time $t_0$ is much less than
one, falling like $1/t_0$.

Since the wavepacket spreads in $\phi$-space, the probability could be
greater if we allow the shifted final state wavepacket's width to vary
from $\sigma_0$. Call the width $\sigma_1$. The probability (before
specifying a scale factor) is then
\begin{equation}
P = \frac{8\pi^2\sigma_0\sigma_1}{\sqrt{T_0^2 + 16\pi^4(\sigma_0^2 + \sigma_1^2)^2}}.
\end{equation}
This can be maximized to go like $\sigma_0/\sqrt{T_0}$, but only if
$\sigma_1 \propto \sqrt{T_0}$, which does not correspond to a
well-defined classical configuration for large $T_0$; this is a more
precise version of the argument at the beginning of this section.

We also consider a linear potential with slope $-b$. Here we will very
briefly summarize the results, while the details of this calculation
can be found in Appendix \ref{apx:linear}. The field now starts at
rest and accelerates down the potential. The final state we take the
overlap with will also have a velocity. If this velocity exactly
matches the velocity of the wavepacket which was evolved in the
potential, then the probability is the same as above,
eq.~\eqref{eq:constp}. If the velocities do not match, there is an
additional exponential suppression which depends on the slope
$-b$. Any constraint on the final velocity of the wavepacket will then
also constrain the length of the linear potential (i.e.~amount of time
the field evolves in the linear potential).

\section{Discussion and Conclusions}

The ``Emergent Universe" paradigm represents an intriguing effort to
construct a cosmology without a past classical singularity.  In this
paper we have analyzed a version of this model in which a scalar field
evolves in a potential that is flat or has a constant slope, to
approximate the asymptotic behavior of the Emergent Universe
potential. Assuming a Gaussian wavepacket form for the wavefunction of
the homogenous mode of the inflaton, we have derived the evolution of
the wavepacket in these two types of potentials with a fixed
background geometry. We then answered the following question: what is
the probability of a well-defined initial wavepacket evolving into
well-defined state after a time $t_0$? In both cases the probability
is proportional to $1/t_0$ for large $t_0$. The Emergent Universe is
built on an infinite past, and thus this probability goes to zero.

It thus appears inconsistent to have both a well-defined
semi-classical approximation to the field and also have infinite past
nonsingular time. If the field has a well-defined value at any given
time (which might be posed as a boundary condition) then evolving back
in time the wave functional was spread over a range of values. The
field velocity will also always have a spread of different values,
most of which do not balance the negative pressure term. If we were to
then include gravity, at yet earlier times the universe would
presumably be a superposition dominated by expanding (from a
singularity) or contracting states.

We stress that we have analyzed only one version of the Emergent
Universe, with a simplified model. Nonetheless, we believe that the
effect that this analysis points to may be rather generic. For
example, consider alternative theories of gravity. The Emergent
Universe has been studied extensively in theories such as
Ho\v{r}ava-Lifshitz, $f(R)$, Loop Quantum Gravity\footnote{In a
  quantum theory of gravity, the quantum aspect of our analysis may be
  modified. In this case, our framework would take place in a
  classical GR limit of quantum gravity (for instance, the size of the
  initial universe is large) or cosmological setting.}, and others
(see, for instance, \cite{Wu:2009ah, Chattopadhyay:2011fp,
  Mulryne:2005ef,
  delCampo:2007mp,*Mukherjee:2005zt,*Paul:2008id,*Banerjee:2007qi,*Cai:2012yf},
respectively). There have also been several studies of the stability
of the Einstein static universe in alternative theories (see
\cite{Boehmer:2009yz,*Boehmer:2007tr,*Seahra:2009ft,*Parisi:2007kv,*Bohmer:2009fc,*Wu:2011xa,*Boehmer:2010xa,*Goswami:2008fs,*Parisi:2012cg},
for example). However, in our framework we have, in a sense, decoupled
gravity -- it enters only when assessing the affect of the spreading
wave-functional. Even in alternative theories in which the Einstein
static universe is more stable than in standard General Relativity, we
anticipate that once the wavefunctional has spread enough, the
geometry must follow, and the spacetime becomes classically
ill-defined as well as containing portions corresponding to
singularities. Therefore, this seems like a generic (and perhaps
expected, given our construction of the scenario) problem with such an
eternal and precisely tuned inflationary scheme.

To avoid this behavior, the field velocity would have to be stabilized
by some mechanism at the correct value, while still allowing for the
field value to evolve appropriately. The potential would have to
remain constant for a static universe, and thus some sort of
(classical) driving and damping terms seem to be necessary. It would
also still be difficult to arrange the appropriate initial
conditions. It is not immediately obvious how one can successfully
achieve this. Another alternative is perhaps a tunneling scenario (for
instance, \cite{Labrana:2011np}). However, then the universe is
necessarily not eternal.

Models in which the field dynamics and material content are very
different would require separate analysis, but may lead to a similar
basic conclusion.  For example, Graham et al.~\cite{Graham:2011nb}
construct static and oscillating universes with a specific
non-perfect-fluid energy component that are stable against small
perturbations.  However, Mithani \& Vilenkin \cite{Mithani:2011en,
  *Mithani:2012ii} have shown that this model is unstable to decay via
tunneling.

Although we have analyzed only one version of the Emergent Universe,
we would argue that our analysis is pointing to a more general
problem: it is very difficult to devise a system -- especially a
quantum one -- that does nothing ``forever,'' then evolves.  A truly
stationary or periodic quantum state, which would last forever, would
never evolve, whereas one with any instability will not endure for an
indefinite time.  Moreover, the tendency of quantum effects to
destabilize even classically stable configurations suggests that even
if an emergent model were possible, it would have to be posed at the
quantum (and quantum-gravitational) level, largely undermining the
motivation to provide an early state in which quantum gravitational
effects are not crucial.

\begin{acknowledgments}
  \noindent
  The authors would like to thank George Ellis and Roy Maartens for
  useful comments on an earlier version of this manuscript. JK would
  like to thank Shinji Mukohyama for useful discussions and to
  acknowledge the UC Santa Cruz Physics Department and SCIPP for their
  support during recent visits while this work was completed. AA was
  supported by NSF Grant PHY-0757911, by a Foundational Questions in
  Physics and Cosmology grant from the John Templeton Foundation, and
  by a New Frontier in Astronomy and Cosmology Grant \#FP050136. This
  work was supported by the World Premier International Research
  Center Initiative (WPI Initiative), MEXT, Japan.
\end{acknowledgments}

\appendix

\section{Linear Potential Calculation}\label{apx:linear}

Here we analyze a potential with a small slope, so the field rolls
down the potential without a need for any initial push as in Section
\ref{sec:const}. The potential is given by
\begin{equation}
u(\phi,T) = 2\pi^2a^6(T)\left(-b\phi + c\right),
\end{equation}
and the initial condition is similar to the flat potential case,
\begin{equation}
  \psi(\phi, -T_0) =
  \left(\frac{1}{2\pi\sigma_0^2}\right)^{1/4}e^{\frac{-(\phi -
      \phi_0)^2}{4\sigma_0^2}}.
\end{equation}

The wavefunction that solves the Schr\"odinger equation in this
potential with this initial condition is
\begin{widetext}
\begin{align}
\psi(\phi, T) =
  \frac{1}{\left(2\pi\sigma_0^2\right)^{1/4}}&\exp\left\{\frac{-\pi^2}{\sigma(T)}\left(\phi
    - \phi_0 -
    ib\int_{-T_0}^T\sigma(T')a^6(T')\mathrm{d}T'\right)^2\right.\nonumber\\
&\left. -\frac{i}{2}\int_{-T_0}^T\frac{1}{\sigma(T')}\left[1 + 4\pi^2\left(c -
      b\phi_0 - ib^2\int_{-T_0}^{T'}\sigma(T'')a^6(T'')\mathrm{d}T''\right)\sigma(T')a^6(T')\right]\mathrm{d}T'\right\}
\end{align}
\end{widetext}
with the same $\sigma(T)$ as in eq.~\eqref{eq:sigma}.

We compute the same probability as in Section \ref{sec:const}, except
that in this case the wavepacket's center moves as
\begin{equation}
\braket{\phi} = \phi_0 + \frac{b(T_0 + T)^2}{2},
\end{equation}
and the momentum changes as
\begin{equation}
\braket{p} = 2\pi^2a_0^6b(T_0 + T),
\end{equation}
where we again assume a constant scale factor (set to $a_0$). The
uncertainties, $\sigma_\phi$ and $\sigma_p$, are the same as for the
constant potential.

Compared to the flat potential, here the field velocity increases
with time, as
\begin{equation}
\braket{\dot\phi} = b(t_0 + t).
\end{equation}
If one wants the field velocity to remain below some critical value
(e.g.~a slow roll condition), then this constrains both the potential
and the amount of time the field evolves in the potential.

To compute the probability as we did previously, the final state
shifted wavepacket needs an additional phase to account for a change
in the momentum,
\begin{equation}
\exp \left[2\pi^2ixa_0^6b(T_0 + T)\phi\right],
\end{equation}
where $x$ is an arbitrary positive real number scaling the momentum of
the wavepacket. With $\chi \equiv 8\pi^2\sigma_0^2$ as in Section
\ref{sec:const}, we find the probability at $T = 0$ to be
\begin{align}
P = &\frac{\chi}{\sqrt{T_0^2 + \chi^2}}\nonumber\\
&\exp\left[-\pi^2\chi a_0^{12}b^2(x - 1)^2 \frac{T_0^2\left(T_0^2 + \half\chi^2\right)}{T_0^2 + \chi^2}\right].
\end{align}
For large $t_0$ (again $T_0 = t_0/a_0^3 \gg \chi$) the probability is
\begin{equation}
P \approx \frac{a_0^3\chi}{t_0}\exp\left[-\pi^2\chi a_0^{6}b^2(x - 1)^2t_0^2\right],
\end{equation}
which falls exponentially fast, unless $x = 1$ (maximizing the
probability with respect to $x$). In this case the wavepackets have
the same final momentum, and the probability reduces to the result of
the flat potential,\footnote{As a check, setting $b = 0$, so the
  potential is a constant, also reproduces the result of the previous
  section.}
\begin{equation}
P \approx \frac{a_0^3\chi}{t_0}.
\end{equation}
Therefore, at best the linear potential can have the same probability,
proportional to $1/t_0$, as the flat potential.

\renewcommand{\bibsection} {\section*{References}}
\bibliography{quant_instab_eu_refs}

\end{document}